\documentclass[aip,showpacs,superscriptaddress,reprint,twocolumn]{revtex4}%
\usepackage{amsfonts}
\usepackage{amsmath}
\usepackage{amssymb}
\usepackage{graphicx}%
\usepackage{psfrag}
\begin{document}
\title{Transport properties of dense deuterium-tritium plasmas}
\author{Cong Wang}
\affiliation{LCP, Institute of Applied Physics and Computational Mathematics, P.O. Box
8009, Beijing 100088, People's Republic of China}
\author{Yao Long}
\affiliation{LCP, Institute of Applied Physics and Computational
Mathematics, P.O. Box 8009, Beijing 100088, People's Republic of
China}
\author{Xian-Tu He}
\affiliation{LCP, Institute of Applied Physics and Computational Mathematics, P.O. Box
8009, Beijing 100088, People's Republic of China}
\affiliation{Center for Applied Physics and Technology, Peking University, Beijing 100871,
People's Republic of China}
\author{Ping Zhang}
\thanks{Corresponding author; zhang\underline{ }ping@iapcm.ac.cn}
\affiliation{LCP, Institute of Applied Physics and Computational Mathematics, P.O. Box
8009, Beijing 100088, People's Republic of China}
\affiliation{Center for Applied Physics and Technology, Peking University, Beijing 100871,
People's Republic of China}

\begin{abstract}

Consistent descriptions of the equation of states, and information
about transport coefficients of deuterium-tritium mixture are
demonstrated through quantum molecular dynamic (QMD) simulations (up
to a density of 600 g/cm$^{3}$ and a temperature of $10^{4}$ eV).
Diffusion coefficients and viscosity are compared with one component
plasma model in different regimes from the strong coupled to the
kinetic one. Electronic and radiative transport coefficients, which
are compared with models currently used in hydrodynamic simulations
of inertial confinement fusion, are evaluated up to 800 eV. The
Lorentz number is also discussed from the highly degenerate to the
intermediate region.

\end{abstract}

 \maketitle

Inertial confinement fusion (ICF) has been a long-desired goal,
since the implosion of deuterium-tritium (DT) capsule at
reactor-scale facilities is a visible way to generate a virtually
unlimited source of energy \cite{PBX:Atzeni:2004,PBX:Lindl:1998}. In
direct-drive ICF, nominally identical Megajoule-class laser beams
illuminate the frozen DT capsule. To trigger the ignition and
maximize the thermonuclear energy gain, high compression of the DT
fuel and a high temperature of the hot spot should be achieved. In
the process of compression, the capsule accelerates and starts to
converge since the laser beams ablate DT shell's surface. The DT
shell would experience a wide range of thermodynamical conditions at
densities up to several hundreds of gram per cubic centimeters
(g/cm$^{3}$) and temperatures from a few to hundreds of electron
volts (eV).

Understanding and controlling the high pressure behaviors of DT fuel
are of crucial interest for the success of ignition experiments. The
equation of states (EOS) of DT fuel are essential for ICF designs,
because the low adiabat compressibility of the material is dominated
by EOS \cite{PBX:Hu:2010}. During the imploding process, impurities
from ablators or hohlraum are inadvertently mixed into the fuel,
which strongly influence the burn efficiency so as the stabilities
of plasmas and the mixing rules. Thus, the viscosity and diffusion
coefficients of DT mixture are important parameters in hydrodynamic
modeling treating interfaces instabilities \cite{PBX:Milovich:2004},
as well as in further investigating multi-physical effects of high-Z
dopants. Finally, accurate knowledge of the electronic and radiative
transport coefficients in dense DT plasmas are also important for
precisely predicting all kinds of hydrodynamical instabilities grown
at the ablation front, at the fuel-ablator interface or at the hot
spot-fuel interface, and play central roles to describe the
evolution of the central hot spot against the cold fuel
\cite{PBX:Recoules:2009}.

To this end, thermophysical properties of DT mixture are determined
by combinations of first-principles molecular dynamics (FPMD) and
orbital-free molecular dynamics (OFMD)
\cite{PBX:Lorenzen:2009,PBX:Lambert:2006} using ABINIT package
\cite{PBX:abinit}. Considering DT mixture at high densities,
Coulombic pseudopotential with short cutoff radius ($r_{cut}=0.001$
a.u.) has been built to avoid overlap between pseudocores
\cite{PBX:Wang:2011}. As a consequence, large cutoff energy
$E_{cut}=$200 Hartree are used. In this letter, we report QMD
calculations for DT mixtures along the 200 to 600 g/cm$^{3}$
isochore with $\Gamma$ point to represent the Brillouin zone and 128
particles in a cubic cell, where FPMD and OFMD are performed up to
100 and 10000 eV respectively. Dynamic simulations are lasted from
8000 to 200000 steps in the isokinetic ensemble. Due to the
computational limits of finite temperature density-functional theory
approach, the electronic structure is calculated up to 800 eV, and a
$4\times4\times4$ Monkhorst-Pack $k$-point mesh is used. In
characterizing the plasma states, the ionic coupling and degeneracy
parameters have been used. The definition of the former one is
$\Gamma_{ii}=Z^{*2}/(k_{B}Ta)$ (the ratio between the electrostatic
potential and the kinetic energy), and the latter one is
$\theta=T/T_{F}$ (the ratio of temperature to Fermi temperature).
The present studied plasmas undergoes from strongly coupled
($\Gamma_{ii}\sim60$) and degenerated states ($\theta\sim0.01$) to
the kinetic states ($\Gamma_{ii}\sim0.01$ and $\theta\sim20$).

\begin{figure}
\includegraphics[width=0.8\linewidth]{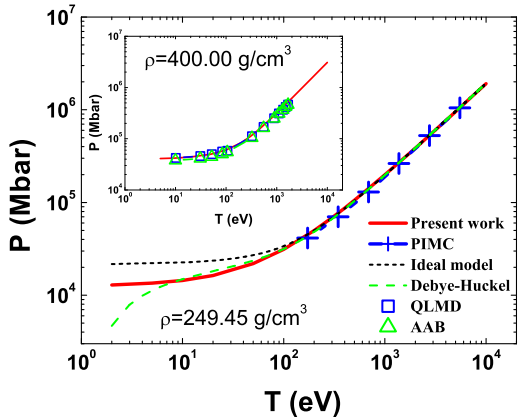}
\caption{(Color online) EOS as a function of temperature at 249.45
g/cm$^{3}$ (Inset is the EOS for $\rho=400$ g/cm$^{3}$). Previous
results, including ideal model and Debye-H\"{u}ckel model, as well
as PIMC, QLMD, and AAB numerical simulations, are also shown for
comparison.} \label{eos}
\end{figure}

Concerning the EOS, the difference between FPMD and OFMD lies within
2\% accuracy below 20 eV, however, as temperature increases, the
difference can be treated as small as negligible. We have
constructed the EOS as polynomial expansions (error within 2 \%) of
the density and temperature ($P=\sum A_{ij}\rho^{i}T^{j}$), which
can be conveniently used in hydrodynamic simulations in ICF or
astrophysics (Table. \ref{coefficient_P}). Figure \ref{eos} shows
the the calculated EOS as a function of temperature at sampled
densities. QMD results are compared with both of the ideal and
Debye-H\"{u}ckel models \cite{PBX:Debye:1923}. The pressure of ideal
model ($P_{id}$) can be viewed as the combination of noninteracting
classical ions and fermionic electrons. The self-consistent solution
of the Poisson equation for a screened charges plasmas leads to the
well known Debye-H\"{u}ckel model, where pressure can be explicitly
expressed as: $P_{D}=P_{id}-\frac{k_{B}T}{24\pi\lambda_{D}^{3}}$.
Here, $\lambda_{D}$ is the Debye length. At a density of 249.45
g/cm$^{3}$, the QMD pressures are 9\% $\sim$ 40\% smaller than those
given by the ideal model, as temperature lies below 100 eV. However,
the Debye-H\"{u}ckel model underestimates the QMD results by 60 \%,
at the lowest temperature (2 eV) considered. For very high
temperatures, beyond $T_{F}$, QMD simulation results and the two
classical methods merge together into the noninteracting classical
ions and electrons (with difference smaller than 1\%). Furthermore,
good agreement in EOS is achieved between the present results and
other simulation methods, such as, Path Integral Monte Carlo (PIMC),
Quantum Langevin molecular dynamics (QLMD), and average atom model
with energy-level broadening (AAB)
\cite{PBX:Hu:2011,PBX:Dai:2010,PBX:Hou:2006}.

\begin{table}[tbh]
\caption{Pressure (kbar) expansion coefficients $A_{ij}$ in terms of
density
(g/cm$^{3}$) and temperature (eV).}%
\centering
\begin{tabular*}{0.8\linewidth}{@{\extracolsep{\fill}}lllr}
\hline\hline $i$ & $A_{i0}$ & $A_{i1}$ & $A_{i2}$\\
\hline
0 & 671.1939 &  816.8465  &  -0.1286   \\
1 & 12645.2534  & 753.1419  &   0.0020 \\
2 & 151.1943  &  -0.0401  &   0.0000 \\\hline\hline
\end{tabular*}
\label{coefficient_P}%
\end{table}

Previous QMD simulations in determining the self-diffusion
coefficients and viscosity were limited at low densities (equivalent
hydrogen mass density up to 8 g/cm$^{3}$) and temperatures (10 eV)
\cite{PBX:Kwon:1994,PBX:Collins:1995,PBX:Clerouin:1997,PBX:Kress:2010}.
The present work substantially extends the study on the transport
coefficients into hot dense regime, and comprehensive comparisons
with one component plasma (OCP) model are described. Here, the
self-diffusion coefficient ($D$) is derived via particle velocity
according to the Green-Kubo relation. The viscosity ($\eta$) is
computed from the autocorrelation function of the five off-diagonal
components of the stress tensor: $P_{xy}$, $P_{yz}$, $P_{zx}$,
$(P_{xx}-P_{yy})/2$, and $(P_{yy}-P_{zz})/2$. In order to reduce the
length of trajectory, empirical functions $C[1-exp(-t/\tau)]$, where
$C$ and $\tau$ are free parameters, have been used to fit the
integrals of the autocorrelation functions \cite{PBX:Kress:2010}.
The fitting procedure is effective in damping the variation and
produces reasonable approximations to $\eta$ (the computed error
lies within 10\%). Due to the fitting procedure and extrapolation to
infinite time, the total error of 20 \% has been estimated. The
uncertainty in self-diffusion coefficient, where an additional
$1/\sqrt{N}$ advantage is secured by particle averages, lies within
2\%.

\begin{figure}
\includegraphics[width=\linewidth]{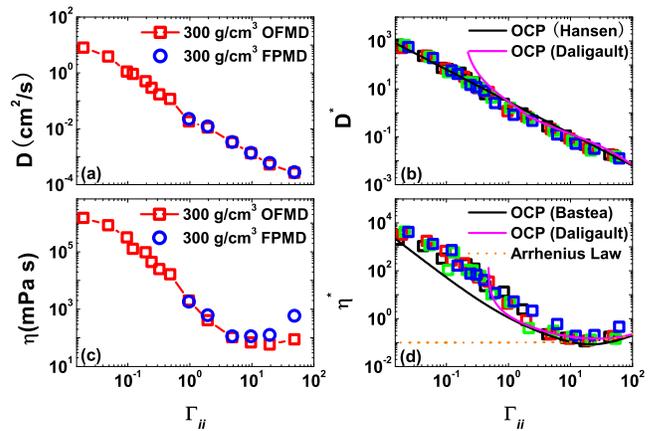}
\caption{(Color online) (a) FPMD and OFMD results for the diffusion
coefficients; (b) Comparison between OFMD simulations and OCP models
for the diffusion coefficients; (c) Viscosity obtained by FPMD and
OFMD; (d) Reduced OFMD viscosities are compared with OCP models. For
the reduced viscosity and diffusion coefficients, the present OFMD
results are denoted by open squares for the density of 200
g/cm$^{3}$ (black), 300 g/cm$^{3}$ (red), 400 g/cm$^{3}$ (green),
and 600 g/cm$^{3}$ (blue).}\label{dif}
\end{figure}

Self-diffusion coefficient and viscosity for DT mixture calculated
by FPMD and OFMD as a function of coupling parameter $\Gamma_{ii}$
at a sampled density of 300 g/cm$^{3}$ are shown in the left panels
in Fig. \ref{dif}. As indicated in the figures, FPMD and OFMD
results for self-diffusion coefficients are in generally good
agreement (error within 5\% or better). As for the viscosity, the
divergence between FPMD and OFMD could reach up to $\sim$ 70\% at
low temperatures, however, as temperature increases, they tend to
merge together. In order to compare these results with OCP
simulations conveniently, the self-diffusion coefficient $D$ and
viscosity $\eta$ are reduced to a dimensionless form:
$D^{*}=D/\omega_{p}a^{2}$, and
$\eta^{*}=\eta/n_{i}M\omega_{p}a^{2}$, where $\omega_{p}=(4\pi
n_{i}/M)^{1/2}Ze$ is the plasma frequency for ions of mass ($M=2.5$
amu.). Memory-function analysis of the velocity autocorrelation
function has been used to obtain the diffusion coefficient of OCP
($D^{*}=2.95\Gamma^{-1.34}$), however, this result is not accurate
at $\Gamma \leq 4$ \cite{PBX:Hansen:1975}. Recently, a more accurate
fit (valid at $0.5 \leq\Gamma\leq 200$) to the OCP simulations has
been provided by Daligault \cite{PBX:Daligault:2006}. In Fig.
\ref{dif} (b), we compare OFMD data with those OCP simulations. The
general feature of our simulation result agrees with that of Hansen
\emph{et al.} \cite{PBX:Hansen:1975}, but visible divergence (3\%
$\sim$ 50 \%) is still observed. The accordance of the Daligault's
fit with the present results is better at low temperatures. As
increasing temperature, the Daligault's OCP fitting overestimates
the OFMD results by $\sim$ 40 \% at $\Gamma_{ii}=0.5$. The present
results for the reduced viscosity are shown in Fig. \ref{dif} (d),
where results from OCP simulations are also provided for comparison.
By performing classical molecular dynamics of OCP, Bastea
\cite{PBX:Bastea:2005} fit $\eta^{*}$ to the form:
$\eta^{*}=0.482\Gamma_{ii}^{-2}+0.629\Gamma_{ii}^{-0.878}+0.00188\Gamma_{ii}$.
The present simulation results for viscosity indicate a similar
behavior between dense plasma and normal fluid, that is, both of
bodily movement of particles and the action of interparticle forces
contribute to the transport of momentum. At large coupling, the
viscosity increases with decreasing temperature (or adding density),
where an Arrhenius-type relation ($\eta^{*}=0.1\times
e^{0.008\Gamma_{ii}}$) is observed. At the intermediate region,
$10\leq\Gamma_{ii}\leq50$, contributions from the two mechanisms
vary with similar magnitude, leading to a shallow minimum. OCP
simulations indicate that the minimum lies at around
$\Gamma_{ii}=25$ \cite{PBX:Daligault:2006}, while our FPMD and OFMD
results suggest the minimum near $\Gamma_{ii}=11$ and
$\Gamma_{ii}=15$, respectively. When $\Gamma_{ii} < 10$, as in a
gas, the bodily movement of particles is predominant and, the
viscosity increases with temperature.

Apart from the EOS and ionic transport coefficients, the electronic
transport coefficients are derived using QMD in the following steps.
Ten snapshots are directly taken from FPMD equilibrium trajectories
up to 100 eV, however, beyond this value, the structure are
extracted from OFMD simulations. Then, the dynamical conductivity
$\sigma(\omega)=\sigma_{1}(\omega)+i\sigma_{2}(\omega)$ is evaluated
through Kubo-Greenwood formula as averages of the selected
configurations. The dc conductivity ($\sigma_{dc}$) follows from the
static limit $\omega\rightarrow0$ of $\sigma_{1}(\omega)$. In the
Chester-Thellung version \cite{PBX:Chester:1961}, the kinetic
coefficients $\mathcal{L}_{ij}$ are described as:
\begin{equation}
\mathcal{L}_{ij}=(-1)^{i+j}\int
d\epsilon\hat{\sigma}(\epsilon)(\epsilon
-\mu)^{(i+j-2)}(-\frac{\partial f(\epsilon)}{\partial\epsilon}%
),\label{coefficient}%
\end{equation}
with $f(\epsilon)$ being the Fermi-Dirac distribution function and
$\mu$ the chemical potential. The electronic thermal conductivity
$K_{e}$ is given by
\begin{equation}
K_{e}=\frac{1}{T}(\mathcal{L}_{22}-\frac{\mathcal{L}_{12}^{2}}{\mathcal{L}_{11}%
}).\label{thermal}%
\end{equation}

\begin{figure}
\includegraphics[width=\linewidth]{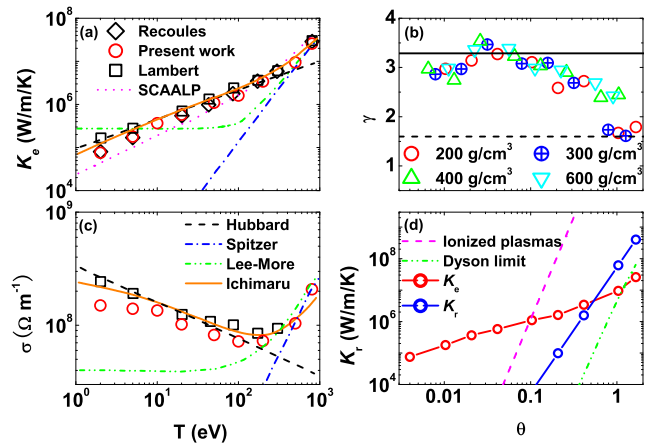}
\caption{(Color online) The present results are compared with those
of classical models and \emph{ab initio} simulations. (a) Electrical
thermal conductivity; (b) Lorentz number; (c) Electrical
conductivity; (d) Radiation thermal conductivity. In panels (a) and
(c) the labels are combined to present the theoretical and QMD
results for $K_{e}$ and $\sigma_{dc}$.}\label{kth}
\end{figure}

The electronic heat conduction as a function of temperature along
the isochore of 200 g/cm$^{3}$ up to 800 eV is shown in Fig.
\ref{kth} (a). Results from theoretical models, which are currently
used in hydrodynamics simulations for ICF, as well as previous
\emph{ab initio} simulations, are provided for comparison. Hubbard
model \cite{PBX:Hubbard:1966} is valid for a highly degenerate
electron system, where the electronic states are treated by using
independent plane waves. The interactions between nuclei are not
screened by electrons, and atomic configurations are assumed to be
Debye- or OCP-like. Unfortunately, thermal conditions in ICF often
enters the partially degenerate regime, which clearly goes beyond
the hypothesis of Hubbard model. Lee-More model \cite{PBX:Lee:1984},
which is not quite accurate in the degenerate state, uses different
formulas for the electron collision time in solid, liquid, and
plasma. The Spitzer model \cite{PBX:Spitzer:1953}, which exhibits a
power law on the temperature ($T^{5/2}$), has been dedicated to
kinetic plasmas. At high temperatures, our QMD results and Lee-More
model merged together into the Spitzer conductivity. From quantum
Boltzmann equation coupled with Ziman theory, Ichimaru
\cite{PBX:Ichimaru:1985} derived hydrogen transport coefficients
restricted to the moderate coupling $\Gamma < 2$. From Fig. 3,
clearly, Ichimaru model results overestimate the present data (up to
$\sim 30\%$) at the strong coupled regime, and then the difference
reduces to 10 \%. Remarkably, our QMD data of the electronic thermal
conductance for DT mixtures shows reasonable consistence with recent
developed self-consistent average-atom model (SCAALP)
\cite{PBX:Faussurier:2010} and precious QMD simulations for hydrogen
\cite{PBX:Recoules:2009,PBX:Lambert:2006} in a wide range.

Beyond the thermal conductance, the dynamic conductivities are also
calculated to examine the Lorenz number defined as:
\begin{equation}
L=\frac{K_{e}}{\sigma_{dc}T}=\gamma\frac{k_{B}^{2}}{e^{2}}.\label{Lorenz}%
\end{equation}
The nature of the screened potential, which is responsible for the
scattering of the electrons, determines $\gamma$. $L$ is constant in
the degenerate and coupled state ($\gamma$ is equal to $\pi^{2}/3$),
reducing to the ideal Sommerfeld number. In the nondegenerate case,
$\gamma$ reaches 1.5966 considering $e$-$e$ collisions. In the
intermediate degenerate region, it is difficult to deduce the
electronic thermal conductivity from electrical conductance by
Wiedemann-Franz law due to the fact that there exists no assumption
on the $\gamma$ value. In Fig. \ref{kth} (b) we show the $\gamma$
value as a function of degeneracy parameter $\theta$ for different
densities. The simulation results indicate that the Lorenz number
vibrates around the Sommerfeld limit in the degenerate region. As
$\theta$ increases, the departure of Lorenz ratio from the ideal
value towards the lower limit is observed.

Results for $\sigma_{dc}$ are presented and compared with those
obtained by classical models or \emph{ab initio} simulations in Fig.
\ref{kth} (c). It has been demonstrated that Hubbard model recovered
the Wiedemann-Franz low in the degenerate limit, and thus the
electrical conductance are computed from the thermal conductance.
For the Spitzer model, the electrical conductivity is obtained by
using the Lorenz number at the lower limit of kinetic matter
including e-e collisions. Our QMD results agree well with Ichimaru
model \cite{PBX:Ichimaru:1985} and data by Lambert \emph{et al.}
\cite{PBX:Lambert:2006} in a wide region between the degenerate
state and the kinetic state.

Another useful coefficient, which plays an important role in the
energy transport formulation of ICF and astrophysics, is the
radiative thermal conductivity ($K_{r}$). As usual, it is connected
to the Rosseland mean path length ($l_{R}=\frac{1}{\rho\kappa_{R}}$
with $\kappa_{R}$ the Rosseland mean opacity - RMO) by the following
relation \cite{PBX:Atzeni:2004}:
\begin{equation}
K_{r}=\frac{16}{3}\sigma_{B}l_{R}(\rho,T)T^{3},\label{radiative}%
\end{equation}
where $\sigma_{B}$ is the Stefan-Boltzmann constant. In particular,
the RMO can be in the form:
\begin{equation}
\frac{1}{\kappa_{R}}=\int_{0}^{\infty}\frac{B'(\omega)}{\alpha(\omega)}d\omega,\label{opacity}%
\end{equation}
where $B'(\omega)$ is the derivative with respect to the temperature
of the normalized Plank's function.
$\alpha(\omega)=\frac{4\pi}{n(\omega)}\sigma_{1}(\omega)$ is the
absorption coefficient with $n(\omega)$ the real part of the index
of refraction. A comparison for $K_{r}$ is made in Fig. \ref{kth}
(d) between our QMD simulations and those given by fully ionized
plasma and Dyson limit. Zeldovich and Raizer have proven that
\cite{PBX:Zeldovich:1998} the RMO of fully ionized plasma can be
obtained as $\kappa_{R}^{ideal}=0.014(Z^{3}/A^{2})\rho
(k_{B}T)^{-7/2}$ cm$^{2}$/g. In this relation, $Z$ is the atomic
number, $A$ is the atomic weight, and $k_{B}T$ is expressed in
kilo-eV. The upper limit of RMO
($\kappa_{R}\leq6.0\times10^{3}Z/(Ak_{B}T)$ cm$^{2}$/g) has been
reported by Bernstein and Dyson via applying Schwartz's inequality
to a particular factoring of the integrand in the definition of RMO
\cite{PBX:Armstrong:1962}. Our QMD simulation results indicate that
$K_{e}$ governs the energy transport process in the fully degenerate
state. As the matter enters the moderate degenerate case, $K_{r}$
rises about four order of magnitude as the temperature increases
from $\sim10^{2}$ eV to $\sim10^{3}$ eV, and becomes dominate when
$\theta$ exceeds $\sim0.5$.

In summary, we have determined the EOS and transport coefficients of
DT plasmas within QMD simulations in the hot dense regime as reached
in future ICF experiments. The wide range EOS has been built from
the coupled to the kinetic regime to describe pressures up to
$10^{6}$ Mbar. A clear chain of simulations in computing
thermophysical properties of hot dense plasma has been demonstrated.
The ionic diffusion coefficient and viscosity have been simulated
and compared with various OCP model simulations. The electronic and
radiative thermal transport coefficients have also been determined,
clearly showing different weights when crossing different
pressure-temperature regimes. The ability to simulate these
parameters in a self-consistent way shown by our results opens a new
way to validate classical theoretical models currently used in
hydrodynamical simulations for ICF and astrophysics.

This work was supported by NSFC under Grants No. 11005012 and No.
51071032, by the National Basic Security Research Program of China,
and by the National High-Tech ICF Committee of China.


\begin{thebibliography}{99}                                                                                               %


\bibitem {PBX:Atzeni:2004}S. Atzeni and J. Meyer-ter-Vehn, \emph{The Physics
of Inertial Fusion: Beam Plasma Interaction, Hydrodynamics, Hot Dense Matter},
International Series of Monographs on Physics (Clarendon Press, Oxford, 2004).

\bibitem {PBX:Lindl:1998}J. D. Lindl, \emph{Inertial Confinement Fusion: The
Quest for Ignition and Energy Gain Using Indirect Drive}, (Springer-Verlag,
New York, 1998).


\bibitem {PBX:Hu:2010}S. X. Hu, B. Militzer, V. N. Goncharov, and S. Skupsky,
Phys. Rev. Lett. \textbf{104} 235003 (2010).

\bibitem {PBX:Milovich:2004}J. L. Milovich, P. Amendt, M. Marinak, and H. Robey,
Phys. Plasmas \textbf{11} 1552 (2004).


\bibitem {PBX:Recoules:2009}V. Recoules, F. Lambert, A. Decoster, B. Canaud,
and J. Cl\'{e}rouin, Phys. Rev. Lett. \textbf{102} 075002 (2009).


\bibitem {PBX:Lorenzen:2009}W. Lorenzen, B. Holst, and R. Redmer, Phys. Rev.
Lett. \textbf{102} 115701 (2009).


\bibitem {PBX:Lambert:2006}F. Lambert, J. Cle¡ärouin, and G. Z\'{e}rah,
Phys. Rev. E \textbf{73} 016403 (2006).


\bibitem {PBX:abinit}Available at http://www.abinit.org.

\bibitem {PBX:Wang:2011}C. Wang, X. T. He, and P. Zhang, Phys. Rev.
Lett. \textbf{106}, 145002 (2011).

\bibitem {PBX:Debye:1923}P. Debye and E. H\"{u}ckel, Phys. Z. \textbf{24}, 185 (1923).


\bibitem {PBX:Hu:2011}S. X. Hu, B. Militzer, V. N. Goncharov, and S. Skupsky, Phys. Rev.
B \textbf{84}, 224109 (2011).

\bibitem {PBX:Dai:2010}J.-Y. Dai, Y. Hou, and J.-M. Yuan, Astrophys. J. \textbf{721}, 1158 (2010).

\bibitem {PBX:Hou:2006}Y. Hou, F. Jin, and J.-M. Yuan, Phys. Plasmas \textbf{13}, 093301 (2006).


\bibitem {PBX:Kwon:1994}I. Kwon, J. D. Kress, and L. A. Collins, Phys. Rev.
B \textbf{50}, 9118 (1994).

\bibitem {PBX:Collins:1995}L. Collins, I. Kwon, J. Kress, N. Troullier, and D. Lynch, Phys. Rev.
E \textbf{52}, 6202 (1995).

\bibitem {PBX:Clerouin:1997}J. G. Cl\'{e}rouin and S. Bernard, Phys. Rev.
E \textbf{56}, 3534 (1997).


\bibitem {PBX:Kress:2010}J. D. Kress, James S. Cohen, D. A. Horner, F. Lambert, and L. A. Collins, Phys. Rev.
E \textbf{82}, 036404 (2010).

\bibitem {PBX:Hansen:1975}J. P. Hansen, I. R. McDonald, and E. L. Pollock, Phys. Rev.
A \textbf{11}, 1025 (1975).

\bibitem {PBX:Daligault:2006}J. Daligault, Phys. Rev. Lett. \textbf{96}, 065003 (2006).

\bibitem {PBX:Bastea:2005}S. Bastea, Phys. Rev. E \textbf{71}, 056405 (2005).

\bibitem {PBX:Chester:1961}G.V. Chester and A. Thellung, Proc. Phys. Soc.
(London) \textbf{77}, 1005 (1961).


\bibitem {PBX:Hubbard:1966}W. B. Hubbard, Astrophys. J. \textbf{146}, 858 (1966).


\bibitem {PBX:Lee:1984}Y. T. Lee and R. M. More, Phys. Fluids \textbf{27}, 1273 (1984).

\bibitem {PBX:Spitzer:1953}L. Spitzer and R. H\"{a}rm, Phys. Rev.
\textbf{89},
977 (1953).

\bibitem {PBX:Ichimaru:1985}S. Ichimaru and S. Tanaka, Phys. Rev.
A \textbf{32}, 1790 (1985).

\bibitem {PBX:Faussurier:2010}G. Faussurier, C. Blancard, P. Coss\'{e}, and P.
Renaudin, Phys. Plasmas \textbf{17}, 052707 (2010).


\bibitem {PBX:Zeldovich:1998}Y. B. Zeldovich and Y. P. Raizer, \emph{Physics of Shock Waves and High Temperature Hydrodynamic Phenomena.}
(Academic Press, New York, 1998).

\bibitem {PBX:Armstrong:1962}B. H. Armstrong, Astrophys. J. \textbf{136}, 309 (1962).


















\end{thebibliography}

\end{document}